\documentclass[runningheads]{llncs}
\usepackage[a4paper, margin=1.3in]{geometry} 
\usepackage[T1]{fontenc}
\usepackage[english]{babel}
\usepackage{amsmath}
\usepackage{amssymb,amsfonts,textcomp}
\usepackage{array}
\usepackage{supertabular}
\usepackage{hhline}
\usepackage{caption}
\usepackage[pdftex]{graphicx}
\usepackage{comment}
\usepackage{subfig}
\usepackage{xcolor}
\usepackage[normalem]{ulem}
\usepackage{booktabs}
\usepackage{mathrsfs}
\usepackage{verbatim}
\usepackage{float}
\usepackage[numbers,sort&compress]{natbib}

\makeatletter

\newcommand\arraybslash{\let\\\@arraycr}
\makeatother
\usepackage{authblk}
\usepackage{graphicx} 

\title{Improving global awareness of linkset predictions using Cross-Attentive Modulation tokens}

\author{Félix Marcoccia\inst{1,2,4} \and
Cédric Adjih \inst{3} \and
Paul Mühlethaler\inst{1}}

\institute{INRIA Paris, Paris, France \and Thales SIX, Gennevilliers, France \and INRIA Saclay, Palaiseau, France \and Sorbonne Université, Paris, France}

\begin{document}
\maketitle

\begin{abstract}
{This work introduces Cross-Attentive Modulation (CAM) tokens, which are tokens whose initial value is learned, gather information through cross-attention, and modulate the nodes and edges accordingly. These tokens are meant to improve the global awareness of link predictions models which, based on graph neural networks, can struggle to capture graph-level features. This lack of ability to feature high level representations is particularly limiting when predicting multiple or entire sets of links. We implement CAM tokens in a simple attention-based link prediction model and in a graph transformer, which we also use in a denoising diffusion framework. A brief introduction to our toy datasets will then be followed by benchmarks which prove that CAM token improve the performance of the model they supplement and outperform a baseline with diverse statistical graph attributes.

}
\end{abstract}

%
\section{Introduction}
\label{sec:introduction}
Learning representations on unstructured or semi-structured data has been one of the major deep learning topics throughout the last decade. From images to graphs by way of text data, multiple architectures have been proposed, and most of them rely on token-level (pixel patches, nodes, edges, words...) feature aggregators that are well fitted to the properties and symmetries of the data modality. Training such aggregators allows to capture robust patterns that grant good generalization properties. For instance, ConvNets \cite{lecun_gradient-based_1998}, Transformers \cite{vaswani_attention_2017} or GNNs \cite{scarselli_graph_2009}, which respectively deal with images, text and graphs, are probably some the most popular and best examples of architectures designed to deal with unstructured or semi-structured data.  Most of these are local or relative, which means that they learn a systematic operator that is applied on each input element independently to allow it to gather some information about the others elements.  This way, each input unit progressively gains information about the other units. To do so, ConvNets mainly learn convolution kernels, while Transformers learn a retrieval-like Query-Key mechanism and GNNs typically learn a message passing scheme. This way of processing inputs differs from other traditional Feedforward Neural Networks such as Multi-layer Perceptrons \cite{Rumelhart1986LearningRB}, which directly combine the inputs according to learned inter-neuron connection patterns. While this straightforward approach can be perfect for many tasks, it is not well suited for most structured data tasks, where the order of the inputs does not indicate any information about the input. When all inputs follow the same set of computations, regardless their position in the input sequence, they are said to go through a permutation-invariant architecture. One major drawback of such architectures is the lack of global pattern aggregation capabilities of the mechanisms that are used. This is especially the fact when dealing with graphs, where nodes struggle to keep track of long range or graph-level dependencies, and tend to be prone to oversmoothing and oversquashing. The domains of generative graph modeling and multi-link prediction can also particularly suffer from this phenomena, as there is no general mechanism that dictates the global distribution of the generated nodes or links, which makes the inference of valid graphs quite challenging. Our goal is to find, for any input set of nodes $V$, a plausible corresponding set of edges $E$. This is often treated as a probabilistic problem that implies estimating the conditional joint distribution of the edges $E$ knowing $V$, which is a classical formulation for conditional generative models. Several variational methods such as \cite{kipf2016variational} or \cite{simonovsky_graphvae_2018} have been proposed. The first one relies on a node-level sampling procedure that assumes nodes' independence, the second one uses a graph level latent space that is then decoded into a graph. The main difficulty of such a method is that it requires complex matching algorithms to compare the reconstructed nodes and edges with the labeled ones. When the decoding is node-conditioned, the conditioning signal also serves as a node identifier (a reconstructed node to which is concatenated an attribute $a_k$ will be compared to the labeled node with attribute $a_k$ if the attributes are unique), which can allow avoiding the matching process. Other methods use generative adversarial formulations to generate realistic graphs. In \cite{decao2022molgan}, the authors use a unconditional Generative Adversarial Network (GAN) helped by objective-driven Reinforcement Learning, an iterative GAN is proposed in \cite{wang2017graphgan} while a temporal sequence GAN is showcased in \cite{lei_gcn-gan_2019}. Flow-based models, such as autoregressive \cite{shi2020graphaf} are also a possible solution for graph generation. More recently, Langevin-esque variational methods have become really popular to generate graphs, in particular Denoising Diffusion Probabilistic Models (DDPM) such as \cite{vignac2023midi} or \cite{vignac2023digress} that respectively provide rotation invariance and discrete, sparsity-encouraging noise schedule. Both use Graph Transformers \cite{dwivedi2021generalization} to denoise at both node and edge levels, as edge attributes are modulated by the nodes' attention scores, which they also modulate, in an interdependent way. Graph Transformers are quite popular and useful to operate at both node and edge levels, and the attention mechanism is particularly well fitted for graph problems, and grants permutation invariance.
As mentioned beforehand, as expressive as it is, the attention mechanism can still struggle to learn graph-level features and is not totally immune to some artifacts, whether by saturating some central nodes with information or creating unwanted high predicted-link-density clusters in areas with an important concentration of nodes. One popular way of improving and providing a form of conditioning to the feature aggregator is to use Feature-wise Linear Modulation (FiLM) \cite{perez2017film} \cite{brockschmidt2020gnnfilm}, which maps a conditional signal to an affine modulation of a feature. For instance, in \cite{vignac2023digress}, the authors use FiLM layers to condition the layers on the time step of the diffusion process, to feature some additional graph-level attributes, and encapsulate some global, statistical information about the nodes and edges. In \cite{marcoccia:hal-04403078}, the authors propose using registers, inspired by \cite{darcet2023vision}, which consists in adding dummy nodes that can be attended to by the true nodes as supplementary information storage. Using attention on the true tokens to gather general information can also so be seen in other domains such as natural language processing with BERT's CLS tokens \cite{devlin2019bert}, in CrossVIT \cite{chen2021crossvit} or in \cite{Wang_2020} for session-based recommendation. While such method is particularly light, it still has limited expressiveness for graphs and does not provide a flexible mechanism to tune the prediction of the links. Our work will then mainly consist in leveraging flexible and expressive information gathering with cross-attention-based register tokens, and in using this information as a conditioning signal for node-level and edge-level modulations.\\
In this work we:
\begin{itemize}
    \item present our CAM tokens' architecture in detail
    \item provide two different architectures that can typically benefit from CAM tokens
    \item present benchmarks that display the merits of our method.
\end{itemize}

\section{Our problem}

Throughout this work, we will tackle a simple toy problem that typically requires both local and global arbitration: given a set of nodes $V$ described by random 2D coordinates, we wish to predict a minimal set of edges $E$, subject to several constraints. One node can not have more than $k$ edges, edges must respect a range constraint, which implies that their length has to be inferior to $d$. The predicted edges have to minimize the number of connected components, hence, if it is possible regarding the range constraint, the graph has to be connected, if not, the number of connected components has to be as small as possible. This can be viewed as a constrained Minimum Spanning Tree problem with all edges' weights being equal to 1. It then implies respecting some local constraints and requires some higher level perception of connectivity of the graph, which is typically what we try to improve, and what node-level message passing schemes can struggle to tackle. The aim of such an approach is not necessarily to get neural networks to perfectly mimic exact optimization results, but to be able to infer plausible graphs that resemble a dataset of graphs that follow the aforementioned constraints.
In order to be valid, the inferred edges $E$ have to:
- be individually valid, present similar properties as the dataset edges and, most importantly, respect the range constraints \\ 
- form coherent local and global patterns, respect the node-wise maximum number of edges, avoid unrealistic amounts of edges, avoid isolated nodes, keep plausible connectivity patterns.\\
The first requirement can be rather easily dealt with by most of the popular model architectures, as it only requires some rather basic information exchange between node pairs. While the main difficulty could reside on the generalization capability of the model, permutation invariance provides strong theoretical robustness to the learned features. The second requirement, especially due to the need of node-level and edge-level permutation invariant operators, is much more difficult to guarantee. Indeed, proposed methods sometimes tend to make "greedy" predictions: if an edge seems plausible in itself, it is inferred, without paying much attention to the context of the other nodes and possible edges. This is precisely the problem we try to tackle here. We also observed that, depending on the spatial distribution of the nodes, some undesirable phenomena could emerge, especially when dealing with high density clusters of nodes, which would often happen to degrade and interfere with the attention-based message passing of nodes even outside the clusters, and also lead to irrelevant link predictions in the clusters.

\section{CAM tokens}
\label{sec:system-model}

\subsection{Our Approach}
As discussed earlier, we extend the principle of attention-based registers, and hereby present CAM tokens. Our method consists in an attentive token that complements a neural network to enhance its global feature aggregation and context awareness capabilities. CAM tokens gather information through cross-attention and learn to modulate the computation of the layers of a model according to the gathered information. 
The value of a CAM token is initialized as
\begin{flalign}
H_{\text{CAM}}^{\ell=0} = \omega_0 \text{, which is a learnable parameter.} 
\end{flalign} \\
Each layer updates the CAM token following
\begin{flalign}
    & H_{\text{CAM}}^\ell = \text{LayerNorm}(\text{CrossAttention}(Q_\text{CAM}^{\ell-1},K_{nodes},V_{nodes}) + H_{\text{CAM}}^{\ell-1}) \\
    & \text{CrossAttention}(Q_{CAM}^{\ell-1},K_{nodes},V_{nodes}) = \frac{Q_{CAM}^{\ell-1} K_{nodes}^T}{\sqrt{d_k}}V_{nodes},
\end{flalign}
with 
\begin{equation}
    Q_{CAM}^\ell = W^QH_{\text{CAM}}^{\ell-1}, K_{nodes} = W^K H, V_{nodes} = W^V H, 
\end{equation} \\
$H$ being the nodes' embeddings. \\ \\
The token value is mapped to a modulation as in \cite{perez2017film}. We choose $\gamma$ and $\beta$ to be simple affine layers. Hence we have 
\begin{flalign}
\gamma^\ell(H_{\text{CAM}}^{\ell}) = W^{\gamma^\ell}H_{\text{CAM}}^{\ell} + b^{\gamma^\ell},\\
\beta^\ell(H_{\text{CAM}}^{\ell}) = W^{\beta^\ell}H_{\text{CAM}}^{\ell}  + b^{\beta^\ell}.
\end{flalign}
A simplified CAM update of the nodes (ignoring the specific architecture and activation function used) is then 
\begin{flalign}
    H^{updated} = \gamma^\ell H + \beta^\ell,
\end{flalign}
with $H$ being the nodes' embeddings.\\ \\
For architectures that feature edge-level representation learning, we can use CAM tokens to modulate also edge features as a function of the information collected from the nodes. We can also use a distinct CAM token that cross-attends to the edges instead of the nodes, and conditions the modulation on both node-level and edge-level CAM tokens. We do it by feeding a small neural network (generally a single linear layer) the concatenation of the node-level CAM token and the edge-level CAM token.\\
A modulation process that incorporates both node and edge features follows 
\begin{equation}
\begin{aligned}
    H_{\text{CAM}}^{\ell} = \text{LayerNorm}( \text{FFN}(\text{CrossAttention}(Q_{\text{CAM}}^{\ell-1},K_{nodes},V_{nodes}),\\ \text{CrossAttention}(Q_{\text{CAM}}^{\ell-1},K_{edges},V_{edges})) + H_{\text{CAM}}^{\ell-1}).
\end{aligned}
\end{equation} \\ \\
\textbf{"Second order" modulation} \\ \\
We also derive a CAM token formulation that allows the context-aware modulation to be a function of the node or edge it is applied to. 
We follow the dot product formulation to introduce an affinity-based modulation that allows nodes or edges to undergo a different modulation depending on their value.  
CAM modulation then follows 
\begin{flalign}
\gamma^\ell(H_{\text{CAM}}^{\ell},H) = W^{\gamma^\ell}(W_{nodes}H (W_\text{CAM}H_{\text{CAM}}^{\ell})^T) + b^{\gamma^\ell},\\
\beta^\ell(H_{\text{CAM}}^{\ell},H) = W^{\beta^\ell}(W_{nodes}H (W_\text{CAM}H_{\text{CAM}}^{\ell})^T) + b^{\beta^\ell}.
\end{flalign}
\begin{figure}[H]
    \centering
    \begin{minipage}{0.46\textwidth}
    \includegraphics[width=\textwidth, height = 0.53\textheight]{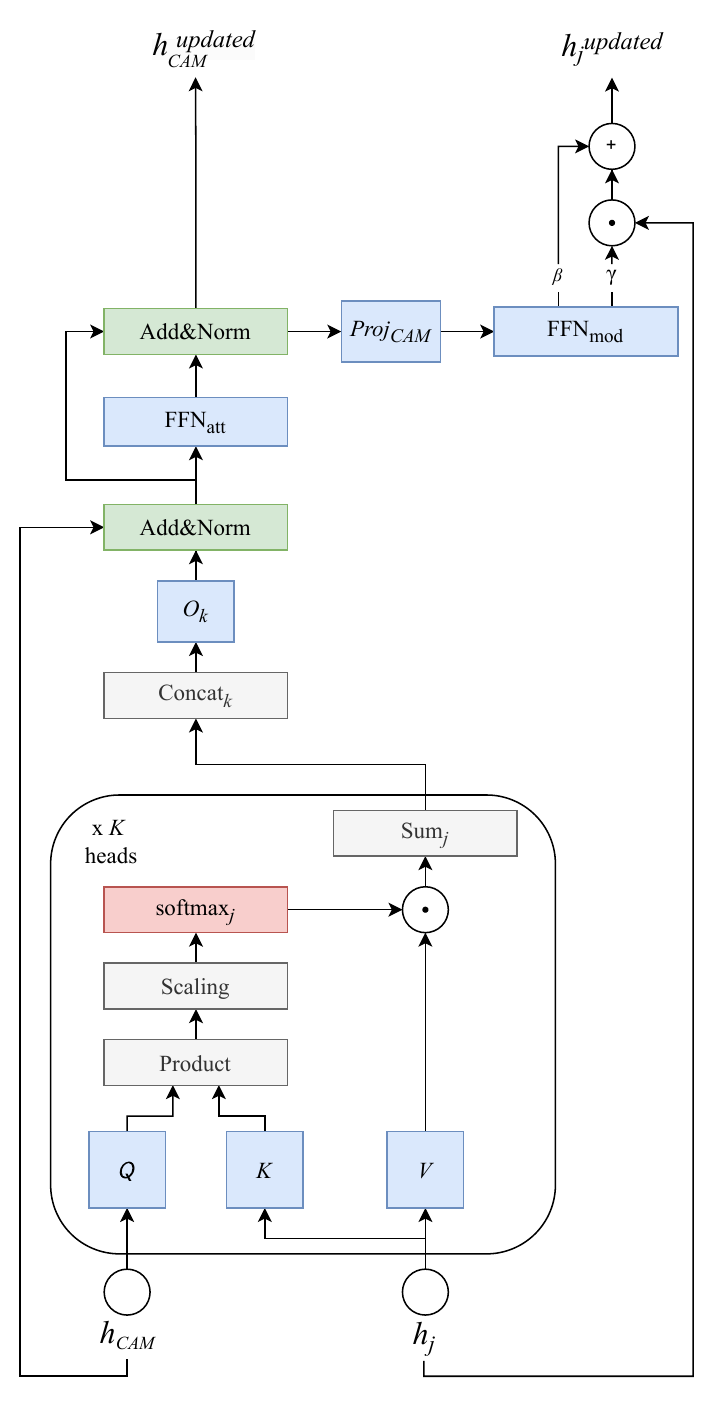}\\
    \par{\centering\textbf{CAM token}\par}
    \label{fig:enter-label}
    \end{minipage}\hfill
    \begin{minipage}{0.46\textwidth}
    \includegraphics[width=\textwidth, height = 0.53\textheight]{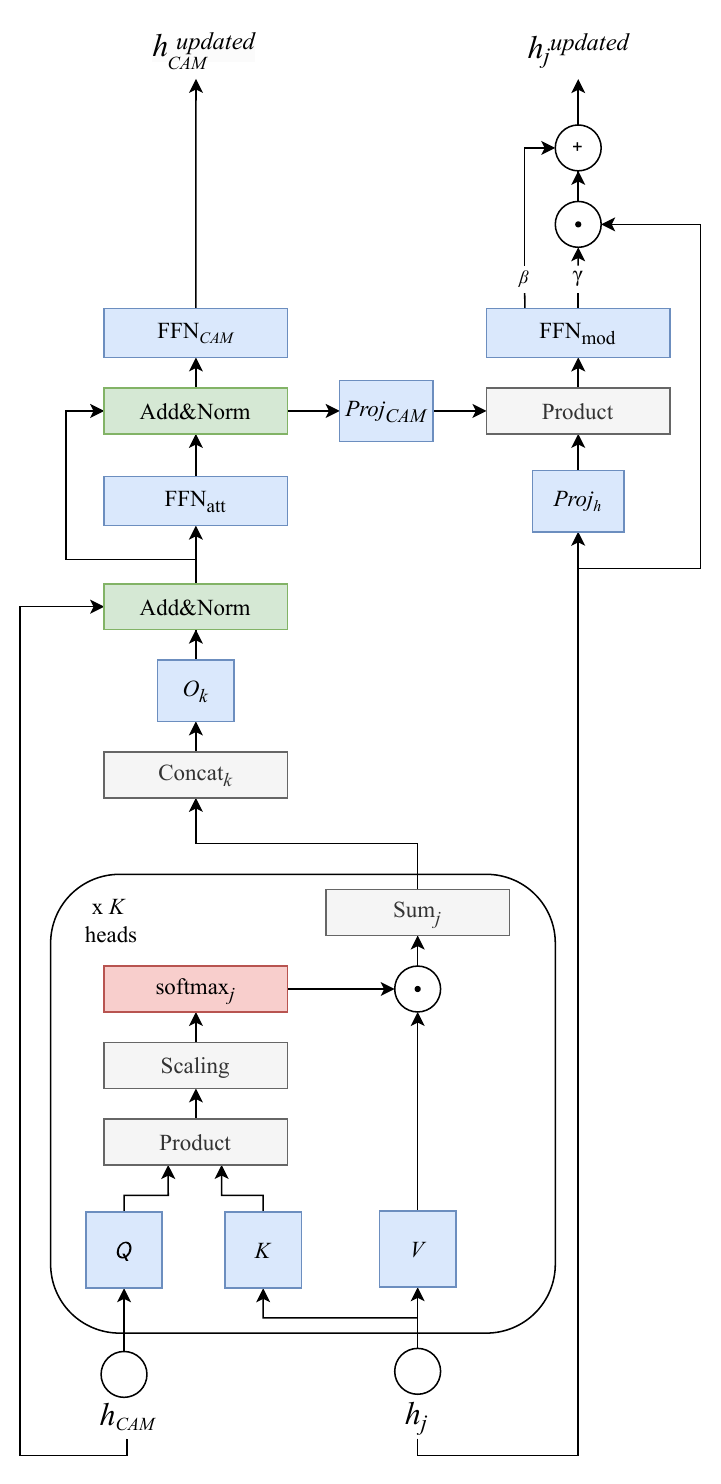}\\
    \par{\centering\textbf{"Second order" CAM token}\par}
    \label{fig:enter-label}
    \par\vspace{1em} 
    \end{minipage} 
    \caption{CAM tokens gather information using cross-attention and modulates the nodes accordingly, allowing to adopt different behaviors depending on the topology of the input nodes. "Second order" CAM tokens gather information using cross-attention and modulates the nodes in function of the dot product of the gathered information and the nodes' embeddings.}
\end{figure}

\subsection{Motivation}
 Typically, we can affiliate CAM tokens, one the one hand, to Registers \cite{darcet2023vision}, cross-attentive modules \cite{chen2021crossvit} or [CLS] \cite{devlin2019bert} attention-based tokens, and, on the other hand, with FiLM layers conditioned on statistical and light handcrafted features, such as in \cite{vignac2023digress}. Our idea is to combine these two expressive and permutation invariant methods, attention and FiLMs, which ultimately reside in conditioning the computation of the embeddings of a token on the value of some intermediate object (the value of the Query-Key product in one case, the value of the FiLM conditioning signal in the other). We seek to follow these guidelines in order to encourage different behaviours depending on the observed features of the system which is being studied. 
 In the first layers, cross-attending the graph allows to gather global information, mainly about the spatial placement of the nodes. Through the layers, nodes and edges have enriched embeddings, which, especially in the last layers, should hold some of the information responsible for the final link predictions. CAM tokens can then, in a first step, modulate the first, basic information gathering layers so that they are well fitted to the particular topology of the set of nodes, and for instance help avoiding oversmoothing in dense areas, or enforce the nodes to attend to isolated nodes. Throughout the layers, CAM tokens can modulate the embeddings to finally guide the prediction towards coherent sets of links. Adding such global mechanism could also help circumventing some of the information bottlenecks brought by oversmoothing and oversquashing, which cause unrealistic, supernumerary link predictions in one case, disconnected nodes in the other, and globally distort the message passing of nodes over the graph. CAM tokens could hence allow anticipating then counterbalancing such behaviours and relieve the message passing mechanism of some of the global aggregation burden.
\section{One shot linkset prediction}
\subsection{Learning setting}
In this section, we can distinguish two approaches that can be used in order to solve our problem. The first one is supervised learning, which learns a deterministic, preferably generalizable mapping between data inputs and outputs. When dealing with high dimensional outputs, especially where different outputs appear to be possible for a single input, or when there are no inputs, a common approach consists in using 
As discussed in \ref{sec:introduction}, most of the methods used to infer sets of links use generative models, which implies approximating the joint probability distribution that we assume our data variables to follow. In our case, the number of possibilities for a set of nodes is rather low, and the information brought by the set of input nodes is strong enough not to necessarily require a generative architecture, and process the problem as multidimensional binary classification. Nevertheless, we derive a typical one-shot generative model that transforms a supervised learning setting into a probabilistic reconstruction problem, a Variational Autoencoder. It follows the same guidelines as \cite{simonovsky_graphvae_2018}, which is particularly practical when dealing with node-conditioned generation. As the encoder, we can use any type of graph-compatible neural network (we use a Graph Transformer), followed by a final pooling layer (similar as in \cite{simonovsky_graphvae_2018} or using a cross-attentive scheme, both presented similar results). The training consists in maximizing
\begin{equation}
\mathcal{L}(\theta, \phi; \mathbf{G}) = \mathbb{E}_{q_{\phi}(\mathbf{z}|\mathbf{G})}\left[ \log p_{\theta}(\mathbf{E}|\mathbf{V},\mathbf{z}) \right]  - D_{KL}\left( q_{\phi}(\mathbf{z}|\mathbf{G})||p(\mathbf{z|\mathbf{G}}) \right),
\end{equation}
where $q_{\phi}$ represents the approximated posterior latent distribution,\\
$p_{\theta}$ represents the approximated likelihood of the data, \\ 
$p(z)$ represents the prior distribution of the latents, we assume it to follow a multivariate Gaussian distribution.  \\ 
At inference time, we sample a $z$ and concatenate it to the desired nodes' coordinates. The decoder then follows
\begin{equation}
decoder(\mathbf{V}) \ \mathtt{\sim} \ p_{\theta}(\mathbf{E}|\mathbf{V},\mathbf{z}).
\end{equation}
\\
In practice, the decoder does not make great use of the encoded vector, and tends to mostly process nodes positions. Different sampled latent vectors still show to result in different predicted graphs, showing that the encoder still holds some useful information and allows for continuous generation.  \\ 
In the following section, we detail different architectures for the decoder, since it is the part of the network that is in charge of computing the links.
\subsection{Example models}
We implement CAM tokens on several permutation invariant architectures that naturally allow to make link predictions. We choose attention-based architectures since they have proved to be particularly expressive for graph problems. \\ \\
\textbf{Attention score model}\\ \\
We implement a small yet efficient model simply composed of a few attention layers (in most experiments we use two) and whose the last attention layers' attention scores are used as thresholded predictions.
An attention layer follows 
\begin{equation}
H^{\ell+1} = \text{Concat}_i\left(\text{softmax}\left(\frac{W^Q_iH^\ell (W^K_iH^\ell)^T}{\sqrt{d_k}}\right)W^V_iH^\ell \right) W^{O},
\end{equation}  \\
where $H^\ell$ represents the embeddings of the nodes at layer $\ell$. $W^Q_i,W^K_i$ and $W^V_i$ are learnable matrices corresponding to the $i$-th attention head.\\
This architecture does not feature edge-level representation learning, but we find attention-score-based prediction to be well suited for our task. Many GNN-based architectures or node-level message passing models rely on inner products to make the link predictions, according to the pair-wise node similarities. A major drawback is that, for such architectures, it is challenging to capture the interdependence of the predicted links: for instance, it is hard for such similarity-based prediction models to respect constraints on the number of predicted links per node. Attention-score-based prediction naturally allows each node to allocate a score to each candidate node, with a total "budget" of 1. We also tried implementing a version with unnormalized attention scores, which tended to perform a bit worse. The set of edges predicted by the last layer is then
\begin{equation}
E^{pred} = \sum_{i=1}^M \text{softmax}\left(\frac{W^Q_iH^{\ell ast} (W^K_iH^{\ell ast})^T}{\sqrt{d_k}}\right)/M,
\end{equation}
with $M$ being the number of attention heads.\\ \\
\textbf{Graph Transformer}\\ \\
We implement a Graph Transformer \cite{dwivedi2021generalization}, with long residual connections. It features interdependent node-level and edge-level representations. It is much deeper than the previous model as it is generally implemented using more than 5 Graph Transformer blocks. A Graph Tranformer layer $\ell$ updates the embeddings following
\begin{equation}
H^{\ell+1} = \text{Concat}_i \Bigg(\text{softmax}\left(\frac{W^Q_iH^\ell (W^K_iH^\ell)^T}{\sqrt{d_k}}W^E_iE^\ell\right)W^V_iH^\ell \Bigg)W^{O_h},
\end{equation} where once again, all heads of attention are then concatenated and combined by a linear layer parametrized by a weight matrix $W_O$.
Edges are updated following \begin{equation}
    E^{\ell+1} =  \text{Concat}_i\left( \frac{W^Q_iH^\ell (W^K_iH^\ell)^T}{\sqrt{d_k}}W_i^EE^\ell \right) W^{O_e}.
\end{equation} \\
\\ 
We use a Graph Transformer since it naturally allows manipulating learnable edge values. It is also one of the most popular architectures for link prediction tasks. The model is much deeper than the previous Attention-based model, by featuring both, we can showcase the effect of CAM tokens on both deep and shallow networks. Since this architecture features edge-level representation learning, we use the node and edge version of CAM tokens mentioned in \ref{sec:system-model}. \\ \\
\subsection{Results}
Our method is implemented using PyTorch. We use AdamW optimizer and generally use a standard binary cross entropy loss function. We use a dataset of multiple instances with randomly generated nodes' positions. It is composed of 190k samples of 16 nodes' positions and the adjacency matrix obtained solving the IP problem with a solver (namely Gurobi). The CAM versions presented below have been implemented using both nodes and edges as features attended by one cross-attentive token, and are both modulated independently. Most of the results are obtained using the "second order" CAM token formulation, as they appeared to be performing slightly better overall. \\
\begin{table}[h]
\centering
\begin{tabular}{c c c c}
\toprule 
\textbf{Model} & \textbf{Accuracy on dataset} & \textbf{Test accuracy} & \textbf{Variance} $\uparrow$\\
\midrule
Att  & 92.49 \% & 85.59 \% & 0.04 \\
\textbf{Att-Stats} & 93.57 \% & 86.88 \% & \textbf{0.06} \\
Att-Reg & 92.55 \% & 85.47 \% & 0.04 \\
\textbf{Att-CAM} & \textbf{93.64 \%} & \textbf{86.90 \%} & \textbf{0.06} \\ 
\textbf{Att-CAM+Stats} & 93.64 \% & 86.83 \% & \textbf{0.06} \\
\bottomrule \\
\end{tabular}
\caption{Performance of the attention model on dataset and unseen graphs with different context awareness add-ons, CAM tokens prove to allow for more accurate and confident predictions.}
\label{tabmilpatt}
\end{table} 
\begin{table}[h]
\centering
\begin{tabular}{c c c c}
\toprule 
\textbf{Model} & \textbf{Accuracy on dataset} & \textbf{Test accuracy} & \textbf{Variance} $\uparrow$\\
\midrule
GT & 94.73 \% & 88.39 \% & 0.10 \\
\textbf{GT-Stats} & 95.89 \% & 89.65\% & \textbf{0.13} \\
GT-Reg & 94.80\% & 88.38 \% & 0.10\\
\textbf{GT-CAM} & \textbf{96.13 \%}   & \textbf{89.77 \%} & \textbf{0.13} \\ 
\textbf{GT-CAM+Stats} & 95.63 \%   & 89.59 \% & \textbf{0.13} \\ 
\bottomrule \\
\end{tabular}
\caption{Performance of the Graph Transformer on dataset and unseen graphs with different context awareness add-ons, CAM tokens once again shows great merits at improving the predictions' accuracy and confidence.}
\label{tabmilp:comparison}
\end{table} 
CAM and stats-augmented models perform better their baseline counterpart, for both Attention model and Graph Transformer, while registers do not seem to be a major asset. Dataset graphs are reproduced more accurately with CAM tokens, and generalize well to unseen graphs. The difference between normal and CAM-augmented models is more important for Graph Transformers, which might signify that CAM tokens are more effective with deeper architectures. It also benefits from the edge-level representation of Graph Transformers, allowing to modulate feature computations at both node and edge-level and thus guiding the link prediction more easily. Using both CAM token and statistics do not seem to improve performance in this setting.\\ \\
When dealing with such one-shot link prediction, the main issue is that the models predict link probabilities instead of true binary edge predictions. In the ideal scenario, correct links would be predicted as 1, and the absence of link would be predicted as 0. In the worst, conservative scenario, the models converge to an easy local minimum that corresponds to the model only predicting an intermediate value, corresponding to the mean of the dataset values In practice, the models behave somewhere between the two. None of the models, even when encouraged with a "mean-repulsive" loss term (which penalizes predictions close to the dataset mean), display binary (or really close to it) scores. We still observe that statistical and CAM tokens versions of the model allow for more confident predictions, as the variance of their predictions are higher. We can suppose that, when many individually equivalent links are possible, the CAM token modulation will tend to accentuate the small dissimilarities between them or to bring some bias that allows the link prediction to be less conservative. Such predictions can either be thresholded to produce the link predictions, or serve as a heuristic to assign the links given their probability. Since the models do not fully commit to confident link predictions, it is frequent that, when "hesitating" between two links, the model predicts both with conservative probabilities, and hence do not model well the interdependence between the links in such situations.
\section{State of the art generative method: Probabilistic Denoising }
\textbf{Node-conditioned Denoising Diffusion Probabilistic Model} \\ \\
Recent works have proven Denoising Diffusion Probabilistic Models (DDPMs) to be particularly effective to tackle the task of generating plausible graphs.
In order to approximate the distribution that we assume our data to follow, graph DDPMs propose, in the continuous setting, to search for a $p_{\theta}^t(\mathbf{G^{t-1}}|\mathbf{G^t})$ that approximates a markovian reverse noise process. This, contrarily to directly maximizing the likelihood of $p_{\theta}(\mathbf{E}|V)$, is directly tractable. In our case, we want to compute the edges $E$ given the nodes $V$, hence we apply the denoising process on edges only. We would then wish to find $p_{\theta}^t(\mathbf{E^{t-1}}|\mathbf{E^t},V) = p_{\theta}^t(\mathbf{E^{t-1}}|\mathbf{G^t})$.\\
We rely on discrete denoising with data-aware state transition matrices, as it proved to be the most effective method in \cite{vignac2023digress}. We then directly model $p_{\theta}^t(\mathbf{E}|\mathbf{G^t})$. The noise is applied to uniform state transition matrices between the node categorical attributes. Here we consider a binary 1-dimensional problem, this is why our "transition matrix" is only a scalar $m$ that is correlated with the true distribution of the dataset links. \\ \\ The noise at step $t$ is given by $\bar{Q}^t = \bar{\alpha}^t I + \bar{\beta}^t \mathbf{m}$, where $\bar{\alpha}^t = \cos ({0.5\pi(t/T + s)}/(1 + s))^2$ with a small $s$ and $\bar{\beta}^t = 1 - \bar{\alpha}^t$. \\
The learning procedure solely resides in minimizing $\sum_{ij}l_{BCE}(E,p_{\theta,ij}(G^t))$.\\ \\ Sampling with the DDPM to generate new graphs iteratively follows \\
\begin{equation}
E^{t-1} \ \mathtt{\sim} \ \prod_{ij}  \hat{p}^E_{(\theta,t,ij)}(G^t) \: q(e_{ij}^{t-1} | e_{ij} = e, e_{ij}^{t}),
\end{equation}\\
with $\hat{p_{\theta}}$ being an estimation of $p$ by a neural network parametrized by $\theta$, $q$ representing the distribution of the scheduled noisy transition matrix. \\ \\
This formulation is particularly valuable since the making of intermediate graph predictions at each timestep  $t$ allows each edge and node to adjust its value depending upon the last observed prediction, which can help modeling the interdependence between the edges. \\Since most DDPMs on discrete objects directly model the final, correct graph at each diffusion step, the problem resembles the previous one-shot approaches. Hence we can then make similar assumptions regarding the lack of global learning mechanisms. Once again, we augment a Graph Transformer architecture with CAM tokens to improve global and spatial feature learning capabilities of the model. 
\section{Results}
\label{sec:results} 
The benchmarked models are implemented using the same denoising diffusion framework as in \cite{vignac2023digress}, with the statistical version featuring the same graph-level statistics. We also feature a model using the eigenvalues of the Laplacian of the graph as an input for FiLM layers. We chose to add Laplacian positional encodings \cite{dwivedi2022benchmarking} to the nodes when they showed to be beneficial. \\
\begin{table}[h]
\centering
\begin{tabular}{l l l l l l l }
\toprule 
\textbf{Model} & \textbf{C.C.} $\downarrow$ & \textbf{Isolated nodes} $\downarrow$ & \textbf{Link validity} $\uparrow$ & \textbf{Saturated nodes} $\downarrow$ & \textbf{Links} \\
\midrule
GT & 3.10 & 1.87 \% & 92.73 \% & 1.93 \% & + 7 \%\\
\textbf{GT-CAM} & 2.68
& \textbf{1.59 \%} & 92.70 \% & \textbf{1.17 \%} & \textbf{+ 0 \%}\\
\textbf{GT-Stats} & 2.70  & 1.74 \% & \textbf{95.32 \%} & 1.19 \% & \textbf{+ 0 \%} \\ 
\textbf{GT-CAM+Stats} & \textbf{2.61} & \textbf{1.59 \%} & 92.31 \% & 1.20 \% & \textbf{+ 0 \%}\\
GT-Reg & 3.11 &  1.87 \% & 92.08 \% & 1.95\% & + 7 \%\\
GT + eigenvalues & 3.02 &  1.89 \% & 93.23 \% & 1.43\% & + 3 \%\\
\bottomrule \\
\end{tabular}
\caption{Benchmark of diffusion models with different context awareness add-ons, CAM tokens showcase improved connectivity management.}
\label{tabmilp:comparison}
\end{table} \\
For each of the models, we retained the version that presented the best results and  trade-off between the different evaluated properties."\textbf{C.C.}" stands for "Connected Components" and measures the number of connected components, excluding isolated nodes. It is a crucial feature since we aim, when possible, at inferring connected graphs, which means that the number of connected components is supposed to be 1 for a set of nodes for which it its possible to infer a connected graph. In practice, the test examples feature an average of $\mathtt{\sim}1.2$ connected components. The ability not to disconnect the graph highly depends on the model's ability to interpret the topology and keep track of the global connectivity of the predicted graph, which is one of the main concerns of our work. Other crucial features are "\textbf{Isolated nodes}", which corresponds to the average number of nodes that do not feature any edge, and "\textbf{saturated nodes}", which represents the average number of nodes that feature supernumerary edges. "\textbf{Links}" is computed as the ratio between the average number of predicted links and the average number of links in dataset graphs. This is an important metric because it is a major factor for a graph to be plausible, and, in our experiments, many training instances would result in models predicting unrealistic amount of links, which also makes the comparison of the other criteria less relevant. The register-based and Laplacian eigenvalues-based formulations do not appear to be particularly effective. We observe that CAM tokens and statistical features greatly improve the performance of the diffusion process, as they display better results for both local and global metrics. Although the statistical features seem more effective at helping predicting valid edges, CAM tokens seem to be useful to prevent disconnecting patterns. We also noticed that, to maintain global connectivity, the statistics-aided model seemed to rely more on the Laplacian positional encodings than the CAM token version does. Using both CAM and stats results in behaviours closer to CAM token version's than to the statistical version's. While it  appears to be a trade-off between both regarding some of the metrics, it also seems to improve global connectivity even further that CAM tokens alone.


\section{Conclusion}
\label{sec:conclusion}
In the course of this paper, we introduced and evaluated rigorously the novel mechanism of CAM tokens, a cross-attentive aggregation and modulation approach aimed at improving the prediction accuracy of plausible linksets. Our application of CAM tokens across various permutation invariant architectures, which are typically deficient in global and flexible control capabilities, demonstrated significant effectiveness. The central objective of our research was to adeptly learn and infer graph structures corresponding to datasets composed of minimal connected graphs that adhere
to specific local constraints. \\ Throughout our experiments, particularly in the context of one-shot learning scenarios, our methodology consistently displayed notable enhancements in model accuracy. These improvements were evident when benchmarked against both traditional baseline models and CAM tokens also showed to outperform models augmented with statistical enhancements. Additionally, the deployment of CAM tokens within Denoising Diffusion Probabilistic Models (DDPMs), which focus on the iterative construction of viable graphs, yielded substantial progress, probably even greater than for one-shot predictions. This advancement was not only in comparison to standard baselines but also demonstrated better management of global graph connectivity over models that solely utilize statistical or spectral features.
The integration of handcrafted statistical features with CAM tokens proved to be particularly synergistic in the context of graph denoising diffusion, significantly boosting the capabilities of the augmented models. While our study was demonstrated using a generic example, the implications of our findings are extensive and applicable across a variety of domains. Notably, domains such as molecule generation, which demands intricate understanding of structural and higher-order properties, or more algorithmic graph domains, such as network topology management, where the maintenance of global connectivity is crucial, could greatly benefit from our approach. Thus, our research suggests that the application of CAM tokens could be universally advantageous, offering substantial contributions to the field of context-aware linkset prediction and graph generation.



\bibliography{ref}
\normalsize

\end{document}